\documentclass[usenatbib]{mn2e} \usepackage{psfig,longtable,amsmath}

\title[Orbital period variations of hot-Jupiters] {Orbital period variations of hot-Jupiters caused
  by the Applegate effect}

\author[C.\,A.\ Watson, and T.\,R.\ Marsh]
       {C.\,A.\ Watson,$^1$\thanks{E-mail: c.a.watson@qub.ac.uk} and
         T.\,R.\ Marsh$^2$ \\ $^1$ Astrophysics Research Centre,
         School of Mathematics \& Physics, Queen's University,
         University Road, Belfast BT7 1NN, UK \\ $^2$ Department of
         Physics, University of Warwick, Gibbet Hill Road, Coventry
         CV4 7AL \\}

\begin{document}
\maketitle
 
\begin{abstract}
Several authors have shown that precise measurements of transit time
variations of exoplanets can be sensitive to other planetary bodies,
such as exo-moons. In addition, the transit timing variations of the
exoplanets closest to their host stars can provide tests of tidal
dissipation theory. These studies, however, have not considered the
effect of the host star. There is a large body of observational
evidence that eclipse times of binary stars can vary dramatically due
to variations in the quadrupole moment of the stars driven by stellar
activity. In this paper we investigate and estimate the likely impact
such variations have on the transit times of exoplanets. We find in
several cases that such variations should be detectable.  In
particular, the estimated period changes for WASP-18b are of the same
order as those expected for tidal dissipation, even for relatively low
values of the tidal dissipation parameter. The transit time variations
caused by the Applegate mechanism are also of the correct magnitude
and occur on timescales such that they may be confused with variations
caused by light-time travel effects due to the presence of a
Jupiter-like second planet. Finally, we suggest that transiting
exoplanet systems may provide a clean route (compared to binaries) to
constraining the type of dynamo operating in the host star.

\end{abstract}


\section{Introduction}
\label{sec:intro}

Since the discovery of the first exoplanet around a solar-like star by
\cite{mayor95}, the exoplanet field has bloomed with over 400
currently known. While the majority of exoplanets have been discovered
through radial velocity measurements of the Doppler wobble effect, it
is the systems that exhibit transits that are most highly
prized. These planets are crucial for determining exoplanet bulk
densities (via radii and mass measurements) as well as their
atmospheric properties (via infrared measurements of their day/night
variations and transmission spectroscopy).

Transiting exoplanets, however, also offer the opportunity to detect
other planets within the system since an additional planet may alter
the period of the observed transits. This can occur in two ways.  In
the first case the gravitational influence of the perturbing body can
can alter the orbital period of the transiting exoplanet
directly. This effect is particularly strong for planets in mean
motion resonances and can even allow Earth-massed objects to be
detected, while `exo-moons' orbiting the transiting planet itself also
induce a similar effect (e.g. \citealt{simon07}).  In the second case,
a perturbing mass in a wider orbit can cause the transiting planet /
star system to wobble around the barycentre, again altering the
observed transit times but by changing the light travel-time.

Although searching for transit timing variations (hereafter, TTVs) can
potentially uncover the existence of Earth-mass objects (see
\citealt{gibson10}; \citealt{rabus09}; \citealt{bean09};
\citealt{gibson09}; \citealt{millerricci08}, and references therein
for recent observational studies), there are other effects that can
lead to TTVs. These include the precession of orbits due to general
relativistic effects (\citealt{pal08}), tidal dissipation, torques due
to the spin-induced quadrupole moment of the host star
(\citealt{miralda02}), perturbations of transit times due to star
spots, as well as reorientation of the planetary orbit with respect to
the Earth as a result of proper motion (\citealt{rafikov09}). However,
there is a wealth of observations of many different eclipsing binary
stars (e.g. \citealt{hall80}; \citealt{glownia86}; \citealt{bond88};
\citealt{warner88}; \citealt*{baptista92}; \citealt{echevarria93};
\citealt{wolf93}; \citealt*{baptista00}; \citealt{baptista02};
\citealt{baptista03}; \citealt{borges08}) that have shown
quasi-periodic variations in eclipse times over timescales of years to
decades that are comparable to, or larger than, the effects being
searched for amongst transiting exoplanets.

The favoured explanation for these observed variations in the orbital
periods of eclipsing binary stars is known as the Applegate effect
(\citealt{applegate92}). This mechanism invokes magnetic activity
cycles in the low-mass components of such binaries to redistribute
angular momentum within the interior of the star, thereby changing the
stellar quadrupole moment which leads to changes in the orbital period
of the components. Later, \cite*{lanza98} proposed that the Applegate
mechanism could also be driven by effectively converting rotational
kinetic energy and magnetic energy back and forth. Regardless of the
details of the exact physical mechanism at work, the Applegate effect
should also operate in most exoplanet systems since the host stars are
(by selection) low-mass stars with a convective outer layer which
should exhibit some form of dynamo activity. It will therefore be
important to know the magnitude of the Applegate effect for exoplanet
systems when interpreting any TTVs. In this paper we briefly review
the Applegate mechanism in the next section, before applying the
analysis of \cite{applegate92} to estimate the effects on known
transiting exoplanet systems. Finally, we look at the implications
that the Applegate effect has for TTV work in detecting additional
planets as well as for the measurement of the level of tidal
dissipation in very-hot Jupiter's.

\section{Period changes in binary stars -- the Applegate effect}
\label{sec:applegate}

Many types of binary stars show evidence for changes in their orbital
periods revealed most easily through eclipse times. If a star suddenly
increases its orbital period $P$ by an amount $\Delta P$, then the
eclipses will arrive progressively later and later until, after a time
$T$, they are delayed by an amount $\Delta t = T\Delta P/P$ with
respect to an ephemeris based upon the initial period. These
variations can be tracked through comparison of observed times to
those calculated assuming linear ephemerides via the so-called `$O-C$
diagrams'.

The period changes observed in close binary stars have typical
magnitudes of $\Delta P/P \sim 10^{-5}$ (which can take on either
sign), and thus significant deviations can build up over time. One
famous example is the 2.87 day-period binary Algol, which has
exhibited deviations from linearity of the order of 3 hours over the
200 years it has been followed. Cataclysmic variable stars, detached
white dwarf/main-sequence binaries, RS CVn stars and W UMa stars all
exhibit similar variations on timescales of years to decades. These
variations, which are at best quasi-periodic, are more often than not
too large to be explained by long-term effects such as nuclear
evolution of the stars, mass loss through winds or angular momentum
loss from gravitational radiation or magnetic braking. The latter in
particular struggles when confronted with orbital periods that {\em
  increase} as well as decrease.

\cite{applegate87} realised that period changes in binary systems
without either mass or angular momentum loss could be driven by
variations in the quadrupole moment of one or both stars. Taking just
one star to have a quadrupole moment $Q$ (in the equatorial plane) and
mass $M$ then, as shown in equation 4 of \cite{applegate92}, its
companion orbits in a gravitational potential of the form,

\begin{equation}
\phi(r) = -\frac{GM}{r} - \frac{3}{2} \frac{GQ}{r^3},
\label{eqn:phi}
\end{equation}

\noindent where $r$ is the distance from the star. The orbital speed
is given by $v^2 = rd\phi /dr$ and therefore, from
equation~\ref{eqn:phi}, if $Q$ increases (the star becomes more
oblate) then the gravitational field in the equatorial plane of the
star also increases. In order to balance for the increased gravity,
the companion then requires to increase its centrifugal acceleration
$v^2/r$ at constant angular momentum ($rv$ is constant). Thus $v$ must
increase and $r$ must decrease and hence the orbital period
decreases. The opposite is true if $Q$ decreases. Under this scenario,
\cite{applegate92} showed that the resulting period changes are given
by (their equation 7):

\begin{equation}
\frac{\Delta P}{P} = -9 \left( \frac{R}{a} \right)^2 \frac{\Delta
  Q}{MR^2},
\label{eqn:deltaP}
\end{equation}

\noindent where $R$ is the radius of the star and $a$ is the orbital
separation of the components. 

The common denominator in all the binaries that exhibit period
variations of the type described above is that at least one of the
components is a low-mass star with a convective envelope  capable of
sustaining a magnetic field generating dynamo. Given that the
variation timescales are of the order of years to decades, stellar
activity cycles are the prime candidate for effecting the changes in
the stellar quadrupole moment. Models of the 1980's
(\citealt{applegate87}; \citealt{warner88}) supposed that
dynamo-generated magnetic fields distorted the star from its
equilibrium shape. These were criticised by \cite{marsh90d} since such
distortions leave the pressure and gravity gradients unbalanced (the
star is driven from hydrostatic equilibrium). This unbalanced force
needs to be balanced by the magnetic field, and \cite{marsh90d} showed
that the stars had insufficient luminosity to drive such changes on
the observed timescales.

This issue was quickly resolved by \cite{applegate92} where magnetic
activity was still invoked to drive the $Q$ variations but rather than
forcing the star from hydrostatic equilibrium, the magnetic fields are
supposed to drive angular momentum transfer within the star.  For
instance, if angular momentum is transported from the core to the
envelope of the star, the star will become more oblate overall. This
required much less energy than the earlier models since the star
performs a transition from one state of hydrostatic equilibrium to
another. With this model \cite{applegate92} was able to explain the
observed period changes in binary stars, and the Applegate effect has
continued to survive the test of time apart from a few refinements
(e.g. \citealt{lanza98}).

\section{Estimating the period changes in exoplanet systems}

The same process that occurs in binary stars will also occur in
exoplanet systems in the case where the host star is magnetically
active. Since the majority of exoplanet hosting stars known to date
are lower main-sequence stars with convective outer layers, these
stars should harbour some form of magnetic field generating stellar
dynamo.

For any given energy budget ($\Delta E$), the equations of
\cite{applegate92} allow the magnitude of the period change,

\begin{equation}
\frac{\Delta P}{P},
\end{equation}

\noindent to be calculated numerically which, for the sake of
completeness, we outline here. \cite{applegate92} adopted a simple
stellar model in which a thin outer shell of mass $M_s$ is spun up by
addition of angular momentum $\Delta J$ from the interior of the
star. The energy required to do this is given by his equation 28,

\begin{equation}
\Delta E = \Omega_{dr} \Delta J + \frac{\left(\Delta J
  \right)^2}{2I_{eff}},
\label{eqn:deltaE}
\end{equation}

\noindent where $\Omega_{dr}$ = $\Omega_s - \Omega_*$ is the angular
velocity of differential rotation between the outer shell ($\Omega_s$)
and the stellar interior ($\Omega_*$). $I_{eff}$ is the effective
moment of inertia given by,

\begin{equation}
I_{eff} = \frac{I_s I_*}{I_s + I_*}
\end{equation}

\noindent where $I_s$ and $I_*$ are the moments of inertia of the
outer shell and the stellar interior, respectively. Typically, for an
outer shell mass of $M_s = 0.1M$, $I_s = I_*$ and therefore in
equation~\ref{eqn:deltaE} we can substitute $2I_{eff} = I_s$, where
$I_s = 2/3 M_sR^2$.

The $\Omega_{dr}$ term on the left-hand side of
equation~\ref{eqn:deltaE} is normally small (but see later) and can be
set to zero.  We can therefore re-express equation~\ref{eqn:deltaE}
as,

\begin{equation}
\Delta E = \frac{3 \left( \Delta J \right)^2}{2 M_s R^2}.
\label{eqn:deltaE2}
\end{equation}

In the model outlined in \cite{applegate92}, variations in the
quadrupole moment, $Q$, of the star are driven by the stellar activity
cycle. Magnetic fields are supposed to drive the angular momentum
transfer from within the star. For instance, if angular momentum is
transported from the core of the star to its envelope, the star will
become more oblate and its quadrupole moment will
increase. \cite{applegate92} computes the rate of change of the
stellar quadrupole as a function of the angular momentum transport in
his equation 26,

\begin{equation}
\frac{ \text{d} Q}{ \text{d} J} = \frac{1}{3} \frac{\Omega R^3}{GM}
\label{eqn:deltaQ}
\end{equation}

\noindent where $\Omega$ is the stellar angular rotation velocity.

As outlined in Section~\ref{sec:applegate}, a change in the stellar
quadrupole moment leads to a corresponding change in the orbital
period given by equation~\ref{eqn:deltaP}, where $a$ is the orbital
separation between, in this case, the star and the planet.  We now
have three equations, with equation~\ref{eqn:deltaE2} relating $\Delta
E$ to $\Delta J$, equation~\ref{eqn:deltaQ} relating $\Delta Q$ to
$\Delta J$, and finally equation~\ref{eqn:deltaP} relating $\Delta P$
to $\Delta Q$. Given that stellar activity cycles are rather variable,
there is no well-defined amplitude or period of variation to be
expected from Applegate's models. The one constraint that we can
invoke is a restriction on the energy budget allowed to drive the
quadrupole moment.  With this in mind, we can use the last 3 equations
to relate the (as yet undefined) energy budget to the change in the
orbital period $\Delta P$ giving,

\begin{equation}
\Delta E = \frac{1}{6} \frac{M}{M_s} \frac{G^2M^3a^4}{\Omega^2 R^8}
\left(\frac{\Delta P}{P} \right)^2
\label{eqn:deltaE3}
\end{equation}

\noindent which can be rearranged to give

\begin{equation}
\frac{\Delta P}{P} = \sqrt{6} \left( \frac{M_s}{M}
\right)^{\frac{1}{2}} \frac{\Omega R^4}{GM^{3/2}a^2} (\Delta
E)^{\frac{1}{2}}.
\label{eqn:deltaP2}
\end{equation}

\noindent If we assume that the power available to drive the
quadrupole changes is some fraction, $f$, of the stellar luminosity
then we obtain a total energy budget of $\Delta E = fLT$, where $T$ is
the timescale over which the quadrupole changes occur. Since stellar
activity cycles tend to be of the order of years or decades
(e.g. \citealt{saar99}), there is potentially a large energy budget
available for driving orbital period changes. If this energy is
supplied by the nuclear luminosity of the star with no energy storage
in the convection zone then (\citealt{applegate92}) the star will
exhibit RMS luminosity variations of

\begin{equation}
\Delta L_{rms} = \pi \frac{\Delta E}{T}.
\label{eqn:lrms}
\end{equation}

\noindent We set the available energy budget such that the luminosity
variations never exceed some fraction $\alpha$ of the total stellar
luminosity. This sets our energy budget as

\begin{equation}
\Delta E < \frac{\alpha LT}{\pi}.
\label{eqn:ebudget}
\end{equation}

\noindent While we have used luminosity variations to set our energy
budget we note that such variations will be strict upper
limits. Indeed, it is not clear that any luminosity changes would be
observable. If the thermal timescale of the envelope is much larger
than the timescale of activity cycles then the observed luminosity may
hardly vary. To test this, we have examined the standard solar model
(SSM) used by \cite{boothroyd03} and estimated the thermal timescale
of the convective envelope (the lower boundary of which we have taken
to lie at a radius $R = 0.713863R_{\odot}$). For each shell in the SSM
the thermal energy due to fully ionised hydrogen, helium and
associated free electrons was calculated (hydrogen and helium are
fully ionised except for the very uppermost regions near the
photosphere). From this we determine a thermal timescale of 73,000
years for the Sun's convective envelope. This is likely to be a lower
limit (but a reasonable estimate nonetheless) since we have not
included the thermal energy from metals. We conclude, therefore, that
any luminosity variations on the timescale of 10's or 100's of years
may be damped considerably, and would be potentially unobservable.

Armed with a prescription for defining the energy budget available to
the star to drive $\Delta Q$ (but being aware of the possible reservations
so far described), we can calculate orbital period modulations via
equation~\ref{eqn:deltaP2}. Following \cite{applegate92}, the
amplitude of the orbital period modulation and the amplitude of the
oscillation observed in an $O-C$ diagram are related by

\begin{equation}
\frac{\Delta P}{P} = 2 \pi \frac{O-C}{T}
\label{eqn:oc}
\end{equation}

\noindent This, combined with equation~\ref{eqn:deltaP2} and setting
$\beta = M_s/M$, leads to potential $O-C$ variations in the observed
planetary transit times of the order,

\begin{equation}
\frac {\delta t}{\text{s}} \leq k
\left(\frac{\Omega}{\Omega_{\odot}}\right)
\left(\frac{M}{\text{M}_{\odot}}\right)^{-\frac{3}{2}}
\left(\frac{R}{\text{R}_{\odot}}\right)^{4}
\left(\frac{L}{\text{L}_{\odot}}\right)^{\frac{1}{2}}
\left(\frac{a}{\text{AU}}\right)^{-2}
\left(\frac{T}{\text{yr}}\right)^{\frac{3}{2}}
\label{eqn:final}
\end{equation}

\noindent where $k = 2.172 \times 10^{-4} \alpha^{\frac{1}{2}}
\beta^{\frac{1}{2}}$.

\section{Results}
\label{sec:results}

\cite{applegate92} showed that the period changes observed in
binaries, with just one exception, can be explained assuming outer
shell masses of $M_s = 0.1M$, and that a total energy budget of the
order of 10\% of the stellar luminosity is available to drive the
quadrupole changes. In calculating the orbital period changes in this
work we have also adopted $M_s = 0.1M$ which sets $\beta = 0.1$ in
equation~\ref{eqn:final}.  We have set $\alpha$ = 0.1, constraining
the total energy budget such that any RMS luminosity variations are
less than 10 per cent of the total stellar luminosity (but note our
previous comments that such luminosity variations are likely to be
heavily damped). From equation~\ref{eqn:ebudget}, this is a factor of
$\pi$ lower than the energy budgets required by \cite{applegate92}. We
should also add that the prescription by \cite{lanza98} is more energy
efficient than that proposed by \cite{applegate92}.  The refinements
by \cite{lanza98} would, therefore, be able to drive larger variations
than those predicted here. Indeed, quasi-periodic variations larger
than those that the \cite{applegate92} prescription could sustain may
have been observed in a few binary systems (see, for example,
\citealt{brinkworth06}; \citealt{lanza05} and references therein).

Taking our conservative energy budget, Table~\ref{tab:trans} lists the
expected O--C variations estimated from equation~\ref{eqn:final} for
the parameters of 59 known transiting exoplanets. These values should
be seen as an order of magnitude estimate of the Applegate effect in
exoplanet systems. Stellar rotation periods are taken from
\cite{watson10} where available, and have been obtained from
measurements of the Ca H \& K emission lines. Other stellar rotation
periods indicated by `$\sim$' are deduced from projected equatorial
rotation velocities, $v \sin i$ and published stellar radii from the
exoplanet encyclopedia. Where only upper limits to $v \sin i$ exist we
can only place lower limits on the stellar rotation periods. Where we
are unable to determine a stellar rotation period we have adopted 30
days in order to derive a reasonable estimate of the amplitude of the
Applegate effect assuming the star has solar-like rotation. The other
parameters, stellar mass, luminosity, orbital period and orbital
separation are taken/derived from the exoplanet encyclopedia.

The remaining variable in equation~\ref{eqn:final} that has not been
discussed is the modulation timescale, $T$, which is related to the
period of the stellar activity cycle. For our Sun, $T$ could be equal
to 11 or 22 years depending on the dynamo at work (see
Section~\ref{sec:dynamo} for a discussion). Since, in general, most of
the host stars of exoplanets are solar-like we have calculated the
deviations due to the Applegate mechanism for both of these
timescales. In addition, we have also calculated the variations over a
50 year modulation period since many of the orbital period variations
in active binaries occur on longer timescales. The $O-C$ variations
derived assuming 11, 22 and 50 year modulation timescales are denoted
$\delta t_{11},  \delta t_{22}$ and $\delta t_{50}$, respectively.

In addition, we have calculated the angular velocity of differential
rotation between the outer shell and the stellar interior as defined
in equation~\ref{eqn:deltaE}. These are expressed as the fractional
change in angular rotation rate $\Delta \Omega / \Omega_{11, 22, 50}$
over the 11, 22 or 50 year modulation period. While the additional
energy required to drive this differential rotation is not included in
equation~\ref{eqn:final} (since it would complicate the equation) we
have included it in an additional code. For the majority of the
systems in Table~\ref{tab:trans} this extra energy requirement has
little effect on the results. Only for those systems with differential
rotation rates above 10$^{-2}$ does equation~\ref{eqn:final} start to
break down. However, these still provide good `order of magnitude'
estimates for the amplitude of the $O-C$ variations that may be driven
by the Applegate mechanism in these systems. For comparison,
\cite{ulrich96} have derived an upper limit of $\delta \Omega/ \Omega
\sim 70 \times 10^{-3}$ over the course of the Sun's 11 year cycle.

Inspection of Table~\ref{tab:trans} shows that the Applegate effect
can produce large, measurable transit time variations in several
systems. These are generally of the order of a few seconds over 11
year modulation periods, up to minutes for systems with 50 year
stellar activity periods. Given these results, we now discuss the
potential implications these findings have on dynamical studies of
exoplanet systems.


\begin{center}

\begin{table*}
\caption[]{Summary of the estimated O--C variations induced by the
  Applegate mechanism for currently known transiting exoplanets. The
  first 5 columns list the system name as well as the host stars'
  rotation period, mass, radius and luminosity. Columns 6 and 7 list
  the planets' orbital period and orbital separation to 4 decimal
  places, respectively. For stars where no constraint could be placed
  on the rotation period we have assumed (for demonstrative purposes)
  a value of 30 days and indicated these entries with an asterisk.
  The next columns indicate the differential surface shear and
  expected  O--C variations for stars with modulation periods of 11,
  22 and 50 years, respectively.  }

\begin{tabular}{lr@{.}lr@{.}lr@{.}lr@{.}lr@{.}lr@{.}lcr@{.}lcr@{.}lcr@{.}l} \hline
Planet & \multicolumn{2}{c}{$P_{rot}$} & \multicolumn{2}{c}{$M_*$} &
\multicolumn{2}{c}{$R_*$} & \multicolumn{2}{c}{$L_*$} &
\multicolumn{2}{c}{$P_{orb}$} & \multicolumn{2}{c}{$a$} &
\multicolumn{1}{c}{$\Delta \Omega/\Omega_{11}$} &
\multicolumn{2}{c}{$\Delta t_{11}$} &  \multicolumn{1}{c}{$\Delta
  \Omega/\Omega_{22}$} & \multicolumn{2}{c}{$\Delta t_{22}$} &
\multicolumn{1}{c}{$\Delta \Omega/\Omega_{50}$} &
\multicolumn{2}{c}{$\Delta t_{50}$} \\

& \multicolumn{2}{c}{(days)} & \multicolumn{2}{c}{(M$_{\odot}$)} &
\multicolumn{2}{c}{(R$_{\odot}$)} & \multicolumn{2}{c}{(L$_{\odot}$)}
& \multicolumn{2}{c}{(days)} & \multicolumn{2}{c}{(AU)} &
\multicolumn{1}{c}{($\times 10^{-4}$)} & \multicolumn{2}{c}{(s)} &
\multicolumn{1}{c}{($\times 10^{-4}$)} & \multicolumn{2}{c}{(s)} &
\multicolumn{1}{c}{($\times 10^{-4}$)} & \multicolumn{2}{c}{(s)}
\\ \hline

WASP-19b      & 10&5 & 0&96 & 0&94 & 0&73 & 0&7888 & 0&0164 & 1.9 &
2&8 & 2.7 & 8&0 & 4.1 & 27&5 \\ Corot-7b      & 23&0 & 0&93 & 0&87 &
0&53 & 0&8536 & 0&0172 & 1.0 & 0&8 & 1.4 & 2&2 & 2.1 & 7&5 \\ WASP-18b
& $\sim$5&6 & 1&25 & 1&22 & 2&2$^*$ & 0&9415 & 0&0203 & 6.1 & 11&5 &
8.6 & 32&4 & 13.8  & 111&0 \\ WASP-12b      & $\sim$36&0 & 1&35 & 1&57
& 3&49 & 1&0914 & 0&0229 & 1.2 & 4&3 & 1.7 & 12&3 & 2.6 & 42&1
\\ OGLE-TR-56b   & 26&3 & 1&17 & 1&32 & 2&1$^*$ & 1&2119 & 0&0225 &
2.0 & 3&0 & 2.8 & 8&4 & 3.6 & 28&7 \\ TrES-3b       & $\sim$27&0 &
0&924 & 0&81 & 0&64 & 1&3062 & 0&0226 & 2.3 & 0&3 & 3.2 & 0&9 & 4.9 &
3&1 \\ WASP-4b       & $\sim$29&0 & 0&9   & 1&15  & 1&09 & 1&3382 &
0&0230 & 2.1 & 1&6 & 3.0 & 4&5 & 4.5 & 15&3 \\ OGLE-TR-113b  & 3&2 &
0&78 & 0&77 & 0&2$^*$ &1&4328 & 0&0229 & 14.9 & 1&5 & 21.1  & 4&4 &
31.8 & 14&9 \\ Corot-1b      & $\sim$10&8 & 0&95  & 1&11  & 1&39 &
1&5090 & 0&0254 & 8.1 & 3&1 & 11.5 & 8&9 & 17.3 & 30&5 \\ WASP-5b
& $\sim$15&7 & 1&021 & 1&08 & 1&26 & 1&6284 & 0&0273 & 6.1 & 1&5 & 8.6
& 4&1 & 13.0 & 14&1 \\ OGLE-TR-132b  & 30&0$^*$ & 1&26  & 1&34  & 2&40
& 1&6899  & 0&0306 & 3.5 & 1&4 & 4.9 & 4&0 & 7.4 & 13&8 \\ Corot-2b
& 4&5 & 0&97  & 0&90 & 0&73 & 1&7430 & 0&0281 & 22.8 & 1&9 & 32.3 &
5&3 & 48.7 & 18&2 \\ WASP-3b       & 5&0 & 1&24  & 1&31  & 2&59 &
1&8468  & 0&0317 & 26.8 & 7&8 & 38.0 & 22&1 & 57.2 & 75&9 \\ WASP-2b
& 30&0$^*$ & 0&84  & 0&83 & 0&46 & 2&1522  & 0&0314 & 4.8 & 0&2 & 6.8
& 0&5 & 10.3 & 1&6 \\ HAT-P-7b      & $\sim$25&0 & 1&47  & 1&84  &
4&95 & 2&2047 & 0&0379 & 6.8 & 4&5 & 9.7 & 12&8 & 14.6 & 43&8 \\ HD
189733b & 13&2 & 0&80 & 0&79 & 0&34 & 2&2186  & 0&0310 & 12.0 & 0&3 &
16.9 & 0&8 & 25.5 & 2&7 \\ WASP-14b      & $\sim$13&4 & 1&319 & 1&30 &
2&66 & 2&2438 & 0&0370 & 9.8 & 1&9 & 13.8 & 5&3 & 20.9 & 18&3
\\ TrES-2b    & 24&8 & 0&98 & 1&00  & 1&05 & 2&4706   & 0&0356 & 9.0 &
0&4 & 12.7 & 1&1 & 19.2 & 3&7 \\ OGLE2-TR-L9b  & $\sim$2&0 & 1&52  &
1&53  & 4&86 & 2&4855 & 0&0308 & 129.4 & 38&1 & 183.1 & 107&8 & 276.0
& 369&4 \\ WASP-1b       & 30&0$^*$ & 1&24  & 1&38 & 2&54 & 2&5200   &
0&0382 & 7.7 & 1&1 & 10.9 & 3&1 & 16.5 & 10&6 \\ XO-2b         &
$>$21&0 & 0&98  & 0&96 & 0&68 & 2&6158  & 0&0369 & 9.9 & $<$0&3 & 14.1
& $<$0&8 & 21.2 & $<$2&8 \\ GJ436b        & 48&0 & 0&452 & 0&46 & 0&04
& 2&6439 & 0&0287 & 3.3 & 0&0 & 4.6 & 0&0 & 7.0 & 0&1 \\ HAT-P-5b
& $\sim$22&7 & 1&16  & 1&17 & 1&55 & 2&7885  & 0&0408 & 12.0 & 0&6 &
16.9 & 1&6 & 25.5 & 5&4 \\ HD 149026b    & $\sim$12&6 & 1&3   & 1&50 &
2&88 & 2&8759 & 0&0431 & 23.1 & 2&8 & 32.6 & 7&9 & 49.1 & 27&0
\\ HAT-P-3b      & $\sim$80&0 & 0&936 & 0&82 & 0&44 & 2&8997  & 0&0389
& 3.1 & 0&0 & 4.4 & 0&1 & 6.6 & 0&3 \\ HAT-P-13b     & $\sim$27&0 &
1&22  & 1&56  & 2&21 & 2&9163   & 0&0426 & 9.6 & 1&5 & 13.6 & 4&3 &
20.5 & 14&7 \\ TrES-1b       & 33&5 & 0&87 & 0&82  & 0&46 & 3&0301  &
0&0393 & 8.6 & 0&1 & 12.1 & 0&2 & 18.3 & 0&8 \\ HAT-P-4b      &
$\sim$14&6 & 1&26  & 1&59  & 2&68 & 3&0565  & 0&0446 & 20.7 & 2&9 &
29.3 & 8&2 & 44.2 & 28&0 \\ HAT-P-8b      & $\sim$6&2 & 1&28  & 1&58
& 3&32 & 3&0763   & 0&0487 & 55.0 & 6&0 & 77.7 & 17&1 & 117.2 & 58&6
\\ WASP-10b      & $>$6&6 & 0&71  & 0&78 & 0&26 & 3&0928 & 0&0371 &
39.6 & $<$0&4 & 55.9 & $<$1&1 & 84.3 & $<$3&9 \\ OGLE-TR-10b   & 15&8
& 1&18 & 1&16 & 1&32 & 3&1013 & 0&0416 & 19.6 & 0&7 & 27.7 & 1&9 &
41.7 & 6&5 \\ WASP-16b      & $\sim$16&0 & 1&022 & 0&95 & 0&76 &
3&1186 & 0&0421 & 19.6 & 0&3 & 27.7 & 0&8 & 41.7 & 2&6 \\ XO-3b
& $\sim$3&8 & 1&213 & 1&38 & 2&91 & 3&1915 & 0&0454 & 106.5 & 6&6 &
150.6 & 18&8 & 227.0 & 64&4 \\ HAT-P-12b     & 30&0$^*$ & 0&73  & 0&70
& 0&21 & 3&2131 & 0&0384 & 9.3 & 0&0 & 13.2 & 0&1 & 19.9 & 0&4
\\ WASP-6b       & $\sim$31&5 & 0&88   & 0&87   & 0&60  & 3&3610  &
0&0421 & 12.0 & 0&1 & 17.0 & 0&3 & 25.6 & 1&1 \\ HD 209458b & 14&9 &
1&01  & 1&15 & 1&47 & 3&5247 & 0&0471 & 31.0 & 0&7 & 43.8 & 2&0 & 66.0
& 6&9 \\ TrES-4b       & $\sim$10&8 & 1&384 & 1&81  & 4&08 & 3&5539  &
0&0509 & 39.2 & 5&4 & 55.4 & 15&3 & 83.5 & 52&3 \\ OGLE-TR-211b  &
30&0$^*$ & 1&33  & 1&64  & 3&87 & 3&6772   & 0&0510 &  16.6 & 1&3 &
23.4 & 3&8 & 35.3 & 13&1 \\ WASP-11b      & $\sim$27&3 & 0&82  & 0&81
& 0&36 & 3&7225  & 0&0439 &  14.5 & 0&1 & 20.5 & 0&2 & 30.9 & 0&8
\\ WASP-17b      & $\sim$7&8 & 1&2   & 1&38  & 3&15 & 3&7354 & 0&0510
&  74.2 & 2&7 & 104.9 & 7&7 & 158.2 & 26&5 \\ WASP-15b      &
$\sim$18&7 & 1&18  & 1&48 & 3&09 & 3&7521 & 0&0542 &  29.1 & 1&3 &
41.2 & 3&8 & 62.1 & 13&1 \\ HAT-P-6b      & $\sim$8&5 & 1&29  & 1&46
& 3&57 & 3&8530  & 0&0524 &  70.3 & 2&9 & 99.4 & 8&1 & 150.0 & 27&6
\\ Lupus-TR-3b   & 30&0$^*$ & 0&87  & 0&82  & 0&38 & 3&9141   & 0&0464
&  14.5 & 0&1 & 20.6 & 0&2 & 31.0 & 0&6 \\ HAT-P-9b      & $\sim$5&5 &
1&28  & 1&32  & 2&55 & 3&9229   & 0&0530 &  105.7 & 2&5 & 149.2 & 7&0
& 225.3 & 23&8 \\ XO-1b         & $>$16&0 & 1&00  & 0&93 & 1&0$^*$  &
3&9415 & 0&0488 &  37.0 & $<$0&2 & 52.3 & $<$0&6 & 78.8 & $<$2&1
\\ OGLE-TR-182b  & 30&0$^*$ & 1&14  & 1&14  & 1&44 & 3&9791    &
0&0510 &  18.4 & 0&2 & 26.0 & 0&7 & 39.2 & 2&3 \\ OGLE-TR-111b  &
30&0$^*$ & 0&82  & 0&83 & 0&41  & 4&0145   & 0&0470 &  16.1 & 0&1 &
22.8 & 0&2 & 34.4 & 0&7 \\ Corot-5b      & $\sim$90&0 & 1&00  & 1&19 &
1&75 & 4&0379 & 0&0495 &  7.1 & 0&1 & 10.1 & 0&4 & 15.2 & 1&3 \\ XO-4b
& $\sim$8&9 & 1&32  & 1&55  & 2&28 & 4&1250   & 0&0555 &  57.3 & 2&4 &
81.0 & 6&7 & 122.1 & 23&0 \\ XO-5b         & $\sim$80&0 & 0&88  & 1&06
& 0&93 & 4&1878 & 0&0487 & 7.5 & 0&1 & 10.6 & 0&2 & 16.0 & 0&9
\\ Corot-3b      & $\sim$4&6 & 1&37  & 1&56  & 4&52 & 4&2568    &
0&0570 & 162 & 6&0 & 229.1 & 16&9 & 345.4 & 57&8 \\ WASP-13b      &
$>$13&8 & 1&03   & 1&34   & 1&86  & 4&3530   & 0&0527 & 48.6 & $<$1&2
& 68.8 & $<$3&5 & 103.7 & $<$12&1 \\ HAT-P-1b   & 19&7 & 1&133 & 1&12
& 1&42 & 4&4653 & 0&0553  & 35.9 & 0&3 & 50.7 & 0&9 & 76.5 & 3&0
\\ HAT-P-11b     & 30&0$^*$ & 0&81  & 0&75  & 0&26 & 4&8878  & 0&0530
& 21.2 & 0&0 & 30.0 & 0&1 & 45.3 & 0&3 \\ WASP-7b       & $\sim$3&7 &
1&28  & 1&24 & 2&30 & 4&9547  & 0&0618 & 254.1 & 2&0 & 359.4 & 5&5 &
541.8 & 19&0 \\ HAT-P-2b   &  4&0 & 1&36  & 1&64  & 3&79 & 5&6335 &
0&0689 & 282.1 & 5&3 & 398.9 & 14&9 & 601.4 & 51&0 \\ Corot-6b      &
30&0$^*$ & 1&055 & 1&03 & 1&30 & 8&8866  & 0&0855 & 100.7 & 0&0 &
142.4 & 0&2 & 214.7 & 0&6 \\ Corot-4b      & 8&9 & 1&1   & 1&15  &
1&75 & 9&2021   & 0&0900 & 369.8 & 0&3 & 523.0 & 0&9 & 788.5 & 3&1
\\ HD 17156b & 22&1 & 1&24 & 1&45 & 2&57 & 21&2169 & 0&1623 & 715.0 &
0&1 & 1011.2 & 0&3 & 1524 & 1&0 \\
\label{tab:trans}

\end{tabular}

\end{table*}

\end{center}


\subsection{Implications for planet detections using TTVs caused
by orbital perturbations by a second planet}

In transiting systems, the presence of a second planet will induce
variations in the transit times (e.g. \citealt{agol05}) and also
durations (e.g. \citealt{kipping09}).  The Applegate effect, however,
will also produce variations in the transit times and durations since
it modulates the orbital period over the course of the stellar
activity cycle. While the variations due to the Applegate mechanism
are at best quasi-periodic, there has been regular confusion as to
whether period variations observed in some binaries are due to the
presence of an orbiting third body or the Applegate mechanism
(e.g. \citealt{soydugan08}). Similarly, in exoplanet systems, it is
possible that TTVs caused by the Applegate mechanism could mimic the
perturbations caused by the presence of a second planet.

To assess the possibility of confusion between TTVs caused by the
Applegate effect and those from a bona-fide additional planet, we need
to consider the form of the TTVs caused by the Applegate mechanism.
First, the magnitude of the TTVs introduced by the Applegate mechanism
drop off as $a^{-2}$ (from equation~\ref{eqn:final}). Thus the
mechanism is only strong for transiting planets with orbital periods
of a few days (which encompasses most transiting systems). Second, the
Applegate effect will cause TTVs that are modulated over the timescale
of the stellar activity cycle, most likely over the course of years to
decades. The question then becomes, are there any possible mechanisms
by which a second planet can modulate the transit times of short
period planets by a few seconds over the course of $\sim$10 years?

\cite{agol05} outline a number of mechanisms by which a second planet
can cause TTVs. Confusion with interior perturbing planets can be
immediately rejected since the modulation of the orbital period will
occur on far too short a timescale, in order to cause TTVs modulated
on decade timescales requires an exterior planet on a relatively wide
orbit. Furthermore, TTVs caused by an additional planet in a
mean-motion resonance orbit generally drive significantly larger
transit-time variations than those possible by the Applegate
mechanism. However, for exterior planets on eccentric orbits with much
larger orbital periods (and not in mean-motion resonance),
\cite{agol05} found that deviations in the transit times accumulate
over the orbital period, $P_{out}$, of the outer planet to give,

\begin{equation}
\delta t = \mu_{out}e_{out}\left(\frac{a_{in}}{a_{out}}\right)^3
P_{out}
\label{eqn:out1}
\end{equation}

\noindent where $\mu_{out}$ is the star/planet mass ratio for the
outer planet, $e_{out}$ is the eccentricity of the outer planet, and
$a_{in}$ and $a_{out}$ are the orbital separations of the inner and
outer planets, respectively.

\begin{figure}
\psfig{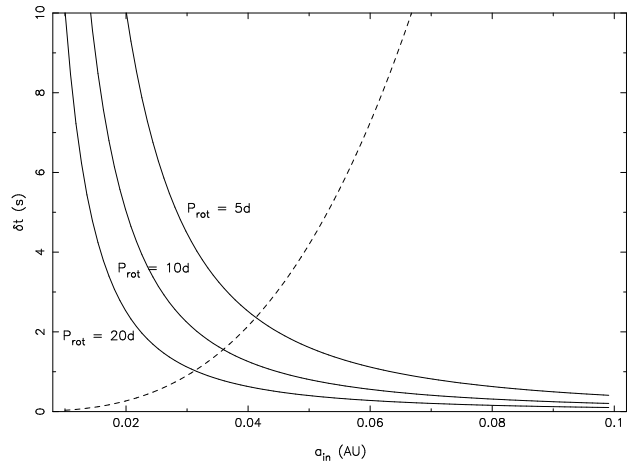}
\caption{The magnitude, $\delta t$, of the transit timing variations
  caused by the Applegate mechanism (solid lines) compared to
  gravitational perturbation by an exterior planet (dashed line) as a
  function of the orbital radius, $a_{in}$, of the interior transiting
  planet. The deviations caused by the Applegate mechanism assume a
  Sun-like host star with rotation rates (indicated on the plot)
  corresponding to a 5, 10 and 20 day stellar rotation period,
  respectively, and a modulation period $P_{mod}$ = 11 yrs. The dashed
  line represents the TTVs caused by a 13M$_J$ mass planet with an 11
  year orbital period ($a_{out}$ = 4.95AU) on a highly eccentric
  ($e_{out}$ = 0.9) orbit.}
\label{fig:ttvplot1}
\end{figure}

Figure~\ref{fig:ttvplot1} shows the magnitude, $\delta t$, of the
transit timing variations caused by the Applegate mechanism (solid
lines) compared to gravitational perturbation by an exterior planet
(dashed line) as a function of the orbital radius, $a_{in}$, of the
interior transiting planet. For the Applegate deviations we have taken
the example of a solar-like host star (M$_*$ = $1M_{\odot}$, R$_*$ =
$1R_{\odot}$, L$_*$ = $1L_{\odot}$) and, as before, set the parameters
$\alpha$ = 0.1 and $\beta$ = 0.1 and assumed a modulation timescale,
$P_{mod}$, of 11 years. We have then plotted three curves
corresponding to stellar rotation rates of 5, 10 and 20 days. To
calculate the maximum deviations expected as a result of gravitational
perturbations by a second planet we have assumed a 13$M_J$ exterior
planet (the accepted upper mass-limit for a planet) in a highly
eccentric ($e_{out}$ = 0.9) orbit (dashed line in
Fig.~\ref{fig:ttvplot1}) with an orbital period of 11 years (matching
the modulation period assumed in the Applegate calculation). This,
therefore, represents the maximum deviations expected from a second
planet. For comparison, a Jovian-mass object in a Jupiter-like orbit
would cause deviations with an amplitude 1/208th that of the object
considered here.

Fig.~\ref{fig:ttvplot1} shows that, for even modest stellar rotation
rates, the Applegate mechanism is likely to dominate over TTVs
 of short period ($a \leq 0.03$AU) transiting systems
around solar-like stars caused by the gravitational perturbations of
a second planet on a long period, non-resonant orbit. For modulation
timescales longer than 11 years (corresponding to longer activity
cycles or longer exterior planet orbital periods) the crossover points
of the curves occurs at larger inner-planet orbital separations,
scaling as $a_{in, xover} \propto P_{mod}^{1/2}$. Since the Applegate
mechanism becomes less efficient as the stellar mass (and luminosity)
decreases, while gravitational perturbations due to a second planet
will increase (due to the $\mu_{out}$ factor in
equation~\ref{eqn:out1}) this situation is likely to reverse for
planetary systems around lower-mass host stars.  Thus, searches for
TTVs around M-dwarf stars may not only be more sensitive to the
presence of planetary companions, but may also be less prone to
'confusion' with TTVs caused by the Applegate effect. (In addition,
M-dwarfs may not have an appreciable dynamo mechanism to drive the
Applegate effect).

We have not considered TTVs for planets in mean-motion resonance
orbits which, under the right conditions, can drive significantly
larger deviations than those described above. The conditions for
possible confusion with the Applegate mechanism (we require long
modulation timescales of years to decades and a short orbital period
for the transiting planet) restricts the perturbing exterior planet to
long orbital periods. The large difference between the orbital periods
of the interior and exterior planets means that there will be no
strong resonance between the components. Mean motion resonances are
unlikely to provide significantly increased gravitational
perturbations since the examples considered here have period ratios
$>$150 (\citealt{beauge05}).

\subsection{Implications for planet detections using TTVs caused
by light travel time effects}

The presence of a second planet in a long period may also cause TTVs
through the variation of the light travel time due to the reflex
motion of the host-star. Indeed, \cite{deeg00} claim tentative
evidence for a Jovian mass object in the eclipsing binary CM Dra by
observing this effect. For an exterior planet of mass $M_p$ on an
orbit with a semi-major axis $a_{out}$, the amplitude of the timing
deviation is,

\begin{equation}
\delta t = \frac{a}{c}\frac{M_p}{M_*} \approx
\left(\frac{M_p}{M_J}\right) \left( \frac{a_{out}}{AU} \right),
\label{lighttime}
\end{equation}

\noindent for a solar-mass star. These periodic variations naturally
occur over a timescale equal to the orbital period of the outer
planet.

For a Jovian-mass object in a Jupiter-like orbit this amounts to a
periodic deviation of $\sim$5 seconds over the course of $\sim$12
years. This is of a similar magnitude and timescale as those
potentially driven by the Applegate effect. There is, therefore,
considerable scope for confusion between TTVs caused by light travel
time delays due to the presence of a second planet and those caused by
the Applegate effect. The only secure means of distinguishing between
these two effect is to look for {\em strictly} periodic variations
since the Applegate mechanism will be quasi-periodic at best.

\subsection{Implications for measuring tidal dissipation}

Recently a subset of exoplanets orbiting extremely close to their host
stars with periods less than 1 day have been discovered. Prime
examples of these are the planets WASP-18b, which was discovered to be
orbiting a 1.24$M_{\odot}$ star every 0.94 days (\citealt{hellier09});
WASP-19b (orbital period = 0.79 days; \citealt{hebb10}); and Corot-7b
(orbital period = 0.85 days; \citealt{leger09}).  These planets are
subject to large tidal forces and can be used as tests of tidal
dissipation theory.  In this scenario, the close proximity of the
exoplanet to the host star raises a tidal bulge on the stellar
surface. This in turn exerts an additional torque which drains angular
momentum from the planetary orbit for systems where the orbital period
is shorter than the stellar rotation period.

In the case of WASP-18, the tidal dissipation is such that assuming a
tidal quality factor $D$ (this is normally denoted $Q$ but we have
changed this to distinguish it from the quadrupole moment used
earlier) found from studies of binaries and the giant planets in our
Solar System means that the planet would be tidally disrupted in
$\sim$1 Myr. Since the system's age is $\sim$5 Gyr, \cite{hellier09}
conclude that either they have caught WASP-18b in an extremely rare
state, or that $D$ is much higher than expected. \cite{hellier09}
quote a period change of -0.00073 (10$^6$/D) s yr$^{-1}$. For our
solar system, $D = 10^5 - 10^6$ leading to orbital period changes of
28 s after 10 yr for $D = 10^6$. Therefore, transit timing studies of
WASP-18b are vitally important since the tidal decay should be
directly measurable in this system.  This would allow our
understanding of tidal dissipation to be tested and would also allow
the stellar interior to be probed since the value of $D$ depends on
how waves generated by tides are dissipated.

The magnitude of the Applegate effect we have estimated for WASP-18
leads to O--C variations of the order of 10's of seconds on decade
timescales -- approximately the same as predicted due to tidal decay.
Such variations could easily either mask, or mimic (briefly), or even
temporarily {\em reverse} any period change caused by tidal decay,
especially if the tidal quality factor is higher than 10$^6$. Thus,
for WASP-18b, caution has to be exercised in interpreting the nature
of any detected transit-time variations, since these will consist of a
quasi-periodic variation on the timescale of the stellar activity
cycle super-imposed on a steady trend of decreasing period due to
orbital decay. Only by observing over the course of an activity cycle
could one begin to reliably distinguish between the two
mechanisms. Thus, WASP-18 is unlikely to reveal its tidal dissipation
history (or future) for several years yet.

\subsection{Implications for stellar dynamo theory}
\label{sec:dynamo}

Finally, we discuss the use of accurate transit timing as a means to
probe the nature of the magnetic field generating dynamo operating in
the host stars' interior. The study of the Applegate mechanism via
eclipse timings of eclipsing binaries is often confused with orbital
period variations caused by angular momentum losses or
exchanges. These include effects due to magnetic braking and mass
transfer. We propose that transiting exoplanet systems, by comparison,
provide much cleaner laboratories in which to study such effects.
Furthermore, unlike in the tidally locked binary systems, exoplanet
systems have the crucial ingredient of a difference between the
stellar rotation period and its orbital period (the orbital period of
the exoplanet). If variations are seen at a level that correlates with
the predictions of Table~\ref{tab:trans} for two stars with similar
orbital periods but different rotation periods, then this would
provide good evidence that the Applegate mechanism is at
work. Naturally, however, one should be aware that gravitational perturbations
caused by an additional long-period planet in the system could
initially be confused with variations due to the Applegate
mechanism. Thus, the quasi-periodic nature of such variations has to
be firmly established before one can be confident that such TTVs can be used
to probe the stellar dynamo.

\cite{lanza98} showed that different dynamo mechanisms result in
different observable manifestations of the Applegate effect. If an
$\alpha \Omega$ dynamo is in operation then the quasi-cyclic orbital
period modulations should occur on the same timescale as the spot
coverage modulation (e.g. over 11 years in the case of our Sun).  In
contrast, if an $\alpha^2 \Omega$ dynamo operates then the observed
orbital period modulation should occur over a timescale twice as long
as the period of the spot modulation ($\sim$ 22 years in the case of
our Sun). Given the large number of transiting exoplanet systems with
host stars (with widely varying fundamental parameters such as age,
rotation rate, masses etc.) that will be discovered by space-missions
such as Kepler and Plato, long-term systematic monitoring for TTVs may
reveal the nature of the dynamo mechanism at work.

\section{Discussion}

We have estimated the likely transit timing variations induced by
changes in the quadrupole moment of the host star in transiting
exoplanet systems driven by the Applegate effect. Depending on the
length of the activity cycle, TTVs of several minutes are plausible
for a number of the currently known transiting exoplanets. While the
timescales and sizes of Applegate driven TTVs are of the wrong
magnitude to be confused with TTVs driven by additional planets in
mean motion resonances, there appears to be much scope for confusion
with TTVs caused by light travel time effects caused by massive,
Jupiter-like planets on wide orbits. 

The magnitude of TTVs driven by the Applegate mechanism also grow as
the star-planet separation decreases (assuming all other factors are
equal). Indeed, for the shortest period transiting exoplanets, such as
WASP-18b, the Applegate mechanism could potentially be mistaken for
orbital period changes due to tidal dissipation. Indeed, the orbital
decay due to tidal dissipation could even be temporarily reversed
since the orbital period variations due to the Applegate effect can
take on either sign. We therefore urge caution when interpreting TTVs,
especially those that appear to be occurring on timescales of years to
decades. In all cases, the clear signature that the Applegate effect
is at work is that the TTVs are quasi-periodic. Only once the strict
periodicity of any TTV has been ascertained should investigators be
confident in their final interpretation.

Finally, an alternative mechanism for driving quasi-periodic stellar
quadrupole variations (and hence orbital period variations) was put
forward by \cite{lanza98}. This rests on the principle that changes in
the azimuthal magnetic field intensity can change the stellar
quadrupole moment by altering the effective centrifugal acceleration,
the stellar dynamo effectively interchanges magnetic energy and
rotational kinetic energy. The main feature of note in this
prescription is that relative changes in the angular velocity required
to drive orbital period changes are a factor of 2 smaller than that
required by \cite{applegate92}.  While it is difficult to assess what
TTVs may be expected from the \cite{lanza98} work, it is quite
possible that non-periodic TTVs with magnitudes exceeding those
outlined in this paper could be observed.  This latter point should
also be taken in light of the fact that we have been very
conservative (in comparison to \citealt{applegate92}) in the energy
budget we have assumed to be available for driving the stellar quadrupole
variations. Indeed, as outlined earlier, quasi-periodic variations
larger than those that the \cite{applegate92} prescription could
sustain may already have been observed in a few binary
systems. Studies of the long-term trends in the transit times of
short-period exoplanets could provide crucial evidence for settling
may of the points above, presenting valuable insights into the working
of stellar dynamos across a broad range of fundamental stellar properties.

\section*{\sc Acknowledgments}

This research has made use of NASA's Astrophysics Data System Bibliographic
Services. The authors also gratefully acknowledge the use of the
Extrasolar Planets Encyclopedia (http://exoplanet.eu/).

\bibliographystyle{mn2e}
\bibliography{abbrev,refs}

\end{document}